\documentclass{ws-p8-50x6-00}
\input epsf
\input{psfig}
\begin{document}

\title{Study of Generalized Parton Distributions\\
 with CLAS at Jefferson Lab}

\author{Volker D. Burkert}

\address{Thomas Jefferson National Accelerator Facility,\\ 
12000 Jefferson Avenue, Newport News, VA 23606, USA}

\maketitle

\abstracts{The program to study the Generalized Parton Distributions in 
deeply exclusive processes with CLAS at Jefferson Lab is discussed.}

\section{Introduction}

In the past nearly five decades of electron scattering, experiments have 
focussed either on the measurements of form factors using 
exclusive processes or on measurements of inclusive processes to
extract deep inelastic structure functions. Elastic processes measure the momentum transfer
dependence of form factors, while the latter ones probe the quark's 
longitudinal 
momentum and helicity distributions in the infinite momentum frame.
Form factors and
deep inelastic structure functions measure two different one-dimensional 
slices of the proton structure. While it is clear that the two pictures 
must be connected, a common framework for the 
interpretation of these data has only been discovered recently with the    
Generalized Parton Distribution (GPD) functions \cite{mueller,xji1,radyush}. 
The GPDs are two-parton correlation functions that encode both
the transverse spatial dependence and the longitudinal 
momentum dependence.  
At the twist-2 level, for each quark species there are two spin-dependent 
GPDs, $\tilde{E}(x,\xi,t),~\tilde{H}(x,\xi,t)$, and 
two spin-independent GPDs, $E(x,\xi,t), H(x,\xi,t)$. The first moments 
of GPDs in $x$ link them to the proton's form factors, while at t=0, the GPDs $H$ 
and $\tilde{H}$ reduce to the quark longitudinal momentum $q(x)$ and 
the helicity distributions $\Delta q(x)$~\cite{belitsky1}, respectively. 
Mapping out the GPDs will allow, for the first time, to construct 
``tomographic'' images of the nucleon's charge and quark helicity 
distributions in transverse impact parameter 
space \cite{burkardt,belitsky2,ralston}. 

The joint probability distribution, represented by the GPDs, contains 
much more of the 
physics of partons than forward parton distributions and form factors.
For example, the spin sum rule~\cite{xji1} makes a direct connection 
between GPDs and the total quark contribution to the nucleon spin:

$$J^q = {1\over 2}\int_{-1}^{1}xdx[H^q(x,\xi,t=0)+E^q(x,\xi,t=0)]$$

The quark helicity in $J^q$ has been measured for the past decade through 
polarized 
deep-inelastic scattering. Therefore, an experimental determination of $J^q$ 
allows a measurement of the quark orbital angular momentum.

\begin{figure}[t]
\vspace{3.8cm}
\centering{\includegraphics{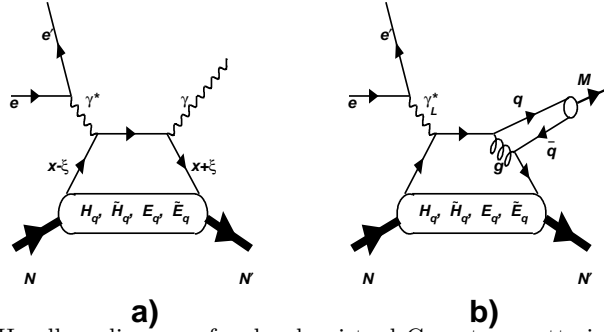}}
\caption{\small{Handbag diagrams for deeply virtual Compton scattering a),
and deeply virtual meson production b).}}
\label{fig:handbag}
\end{figure}

\section{Probing GPDs at Jefferson Lab}

The basis for getting access to GPDs is the ``handbag'' mechanism for 
deeply virtual exclusive process 
shown in Fig.\ref{fig:handbag}. 
The process must be controlled by the pQCD evolution to probe the 
nucleon dynamics. Hard production of high energy photons (Deeply Virtual 
Compton Scattering, DVCS) is likely to
enter the Bjorken regime at relatively low photon virtuality, $Q^2$. 
At the energies currently available at JLab, the DVCS process 
is masked by the more copious production of photons from the Bethe-Heitler (BH)
process. However, using polarized electron beams allows to isolate the
DVCS/BH interference term, which is related to the GPDs 
as:
$$\Delta{\sigma}\sim \sin{\phi}[F_1H(\xi,\xi,t) + k_1(F_1+F_2)
\tilde{H}(\xi,\xi,t)+k_2F_2E(\xi,\xi,t)]~,$$
where $F_1,~F_2$ are the Dirac and Pauli form factors of the nucleon, 
$k_1,~k_2$ are kinematical quantities, and $\phi$ is the angle 
between the $\gamma^*\gamma$ plane and the electron scattering plane.    

\begin{figure}[t]
\vspace{6.2cm}
\centering{\includegraphics{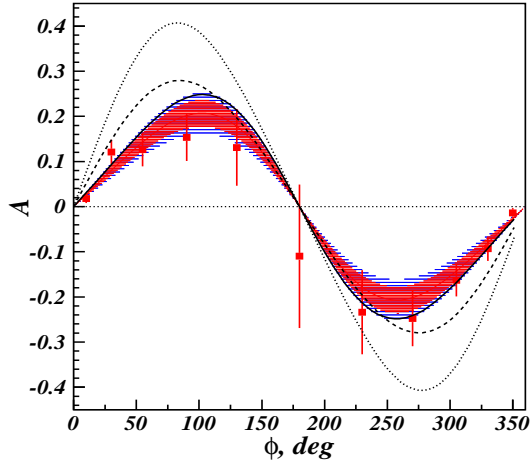}}
\caption{\small{Beam spin asymmetry measured with CLAS. The shaded 
band shows the systematic uncertainties. The curves represent early 
predictions within the GPD framework.}}
\label{fig:clasdvcs}
\end{figure}

The well known BH term is used here to ``boost'' the much smaller DVCS terms
which depends on the unknown GPDs.  
The asymmetry, which is due to the interference, has recently been 
measured~\cite{dvcsclas,dvcshermes}. 
The polarization 
asymmetry is shown in Figure \ref{fig:clasdvcs} compared to various GPD 
predictions~\cite{mvdh,belitsky3}. The CLAS data have been 
predicted within the GPD framework using cross section data from 
HERA~\cite{dvcsh1,dvcszeus}
measured at very small $x_B$ values as input~\cite{freund}. 
The measured asymmetry is in excellent agreement with the 
prediction in LO. 
Other channels are currently under investigation with CLAS, e.g. 
$\gamma^*_{L}p\rightarrow p\rho^0$. For meson production the longitudinal 
component of the photon is of interest. L/T separated cross section data taken at 
$Q^2= 1.5 - 2.3$ GeV$^2$ have been analysed, and are 
approximately described within the GDP formalism, together with data 
from HERMES and Fermilab using a ``frozen'' $\alpha_s$ value~\cite{guidal}.  
The results on both DVCS and deeply virtual 
$\rho$ production allow us to make a rough estimate of the $Q^2$ 
range where the handbag diagram may give the dominant contribution to the reaction, to 
$Q^2 > 1 - 2 $ GeV$^2$ for DVCS/BH, and $Q^2 > 2 - 5$ GeV$^2$ for vector meson 
production.

\section{GPD Program at 6 GeV}

\begin{figure}[t]
\vspace{6.5cm}
\centering{\includegraphics{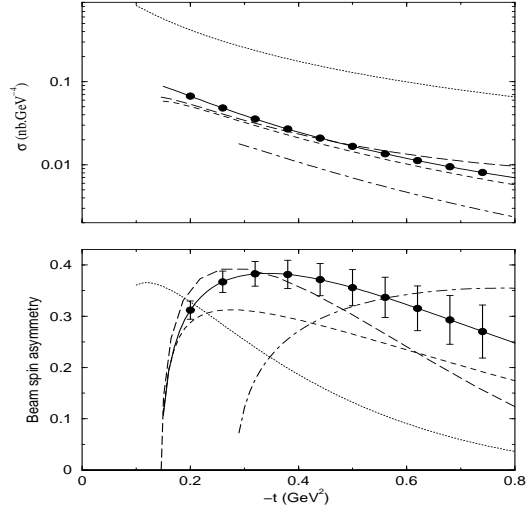}}
\caption{\small{Expected uncertainties for the $t$-dependence 
in experiment E-01-113. Errors are shown for 
$Q^2 = 2$GeV$^2$, $x_B = 0.35$, $\phi=90^{\circ}$. 
The dashed and long-dashed lines are for the same kinematics but different
GPD ingredients. The other lines correspond to different kinematics 
that will be measured simultaneously.}}
\label{fig:e01-113}
\end{figure}

Data at 5.75 GeV have been taken with CLAS to measure vector meson 
and pion production in the deep inelastic region 
and further explore the validity of the GPD framework for meson production~\cite{garcon}. 
Also DVCS/BH asymmetries have been
measured with an order of magnitude improved statistics over the first
measurement at 4.3 GeV. 
Experiments are in preparation at JLab with CLAS and in Hall A to  
measure the DVCS/BH interference in a range $Q^2= 1.5- 3.5$~GeV$^2$ using 
a set of new detectors that will allow a complete measurement of all final 
state 
particles (e, $\gamma$, p)~\cite{hallb,halla}. The $Q^2$ dependence 
will be precisely measured, as will the $\xi$ and $t$-dependences, 
allowing to obtain a first detailed view at the 
kinematical dependences of DVCS, and to test GPD models. 
Figure~\ref{fig:e01-113} shows projected errors for some of the 
kinematics points that will be measured. 

There will also be first results on DVCS/BH asymmetries using a 
longitudinally polarized hydrogen target. While beam spin asymmetries 
probe the GPD $H(\xi,\xi,t)$, longitudinal target asymmetries 
are sensitive to the GPD $\tilde{H}(\xi,\xi,t)$. 

\section{GPD with the CEBAF Upgrade to 12 GeV}

At 11 GeV, the DVCS and BH cross section become comparable in 
size in a broader kinematics domain.
Since the BH cross section is well known, this will allow direct 
extraction of the DVCS 
cross section. While beam asymmetries give access to
the imaginary part of the DVCS amplitude, the DVCS cross section 
determines the $x$-integral,
and contains other combinations of GPDs, therefore 
providing independent information.      

\begin{figure}[t]
\vspace{5.7cm}
\centering{\includegraphics{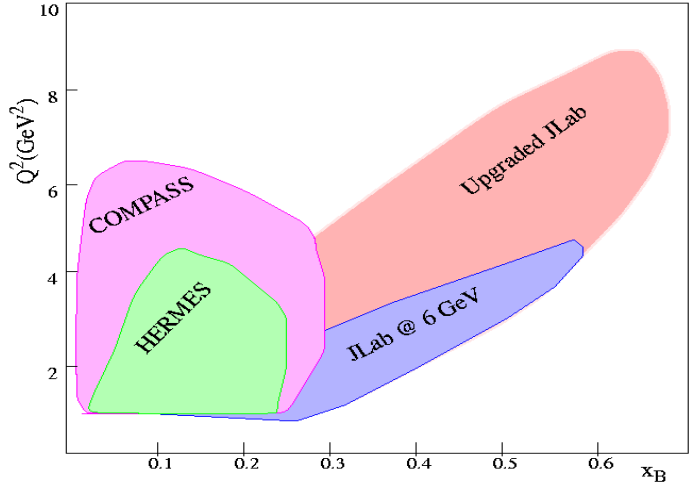}}
\caption{\small{Kinematics for deeply virtual exclusive processes 
for different laboratories, including the proposed energy upgrade 
of CEBAF at JLab.}}
\label{fig:kinematics}
\end{figure}

\begin{figure}[t]
\vspace{6cm}
\centering{\includegraphics{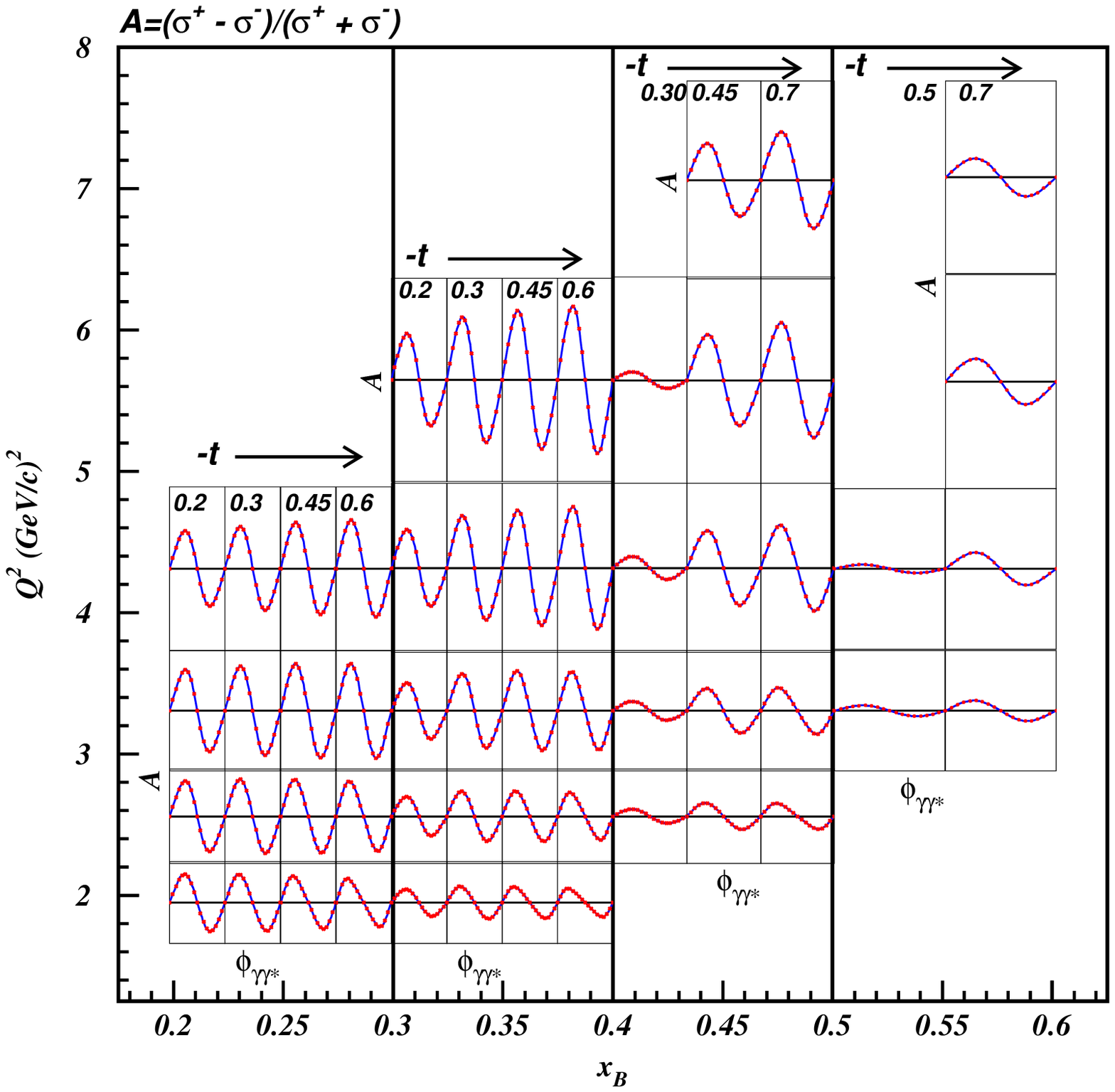}}
\centering{\includegraphics{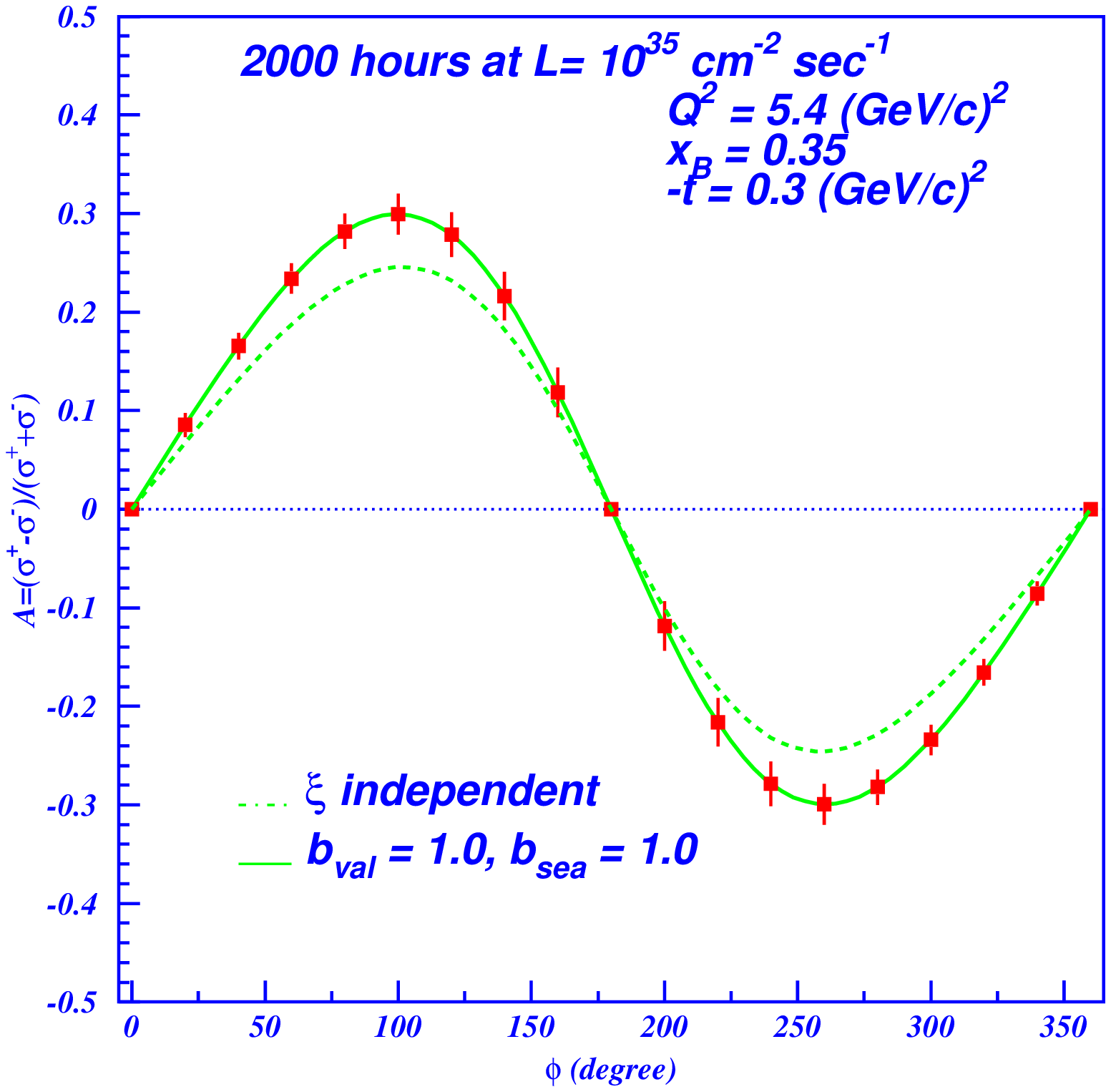}}

\caption{\small{Left panel: Kinematics for DVCS beam asymmetry measurements 
at 11 GeV. Only the bins for lower $t$ values are shown. All bins 
will be measured simultaneously with CLAS$^{++}$. Right panel: 
One of the bins from the left panel is shown with projected error bars. 
The curves represent two models for GPDs.}}
\label{fig:dvcs12gev} 
\end{figure}

The proposed energy upgrade of the JLab accelerator, in conjunction with 
the unprecedented luminosity that will be available, will allow
a much broader kinematic coverage to be accessed in measurements of 
deeply virtual 
exclusive processes. Figure~\ref{fig:kinematics} shows 
the expected coverage in $Q^2$ and $x_B$. Although the energy is lower
than for other experiments, the luminosity will be several 
orders of magnitudes greater for experiments at JLab. 
For the upgraded CLAS (CLAS$^{++}$) an operating luminosity of 
$10^{35}$cm$^{-2}$sec$^{-1}$ is anticipated. Figure~\ref{fig:dvcs12gev} 
shows the projected coverage
of the beam spin asymmetry measurement. 
These measurements will produce high precisison 
DVCS data for $Q^2 = 1.0 - 7.5$GeV$^2$, $x_B = 0.1 - 0.65$, and 
$-(t - t_{min}) <  1.5$ GeV$^2$. 

\section{From Observables to GPDs}

Extracting GPD information from asymmetries and cross section measurements
is not an easy task, and in general may not give unambiguous results. 
However, we can use constraints given by form factors as well as DIS 
experiments, and all deeply exclusive processes may be used in such an 
analysis. In addition, the GPDs are 
strongly constraint by certain polynomial conditions. 
Currently, at least three avenues are being investigated on how to obtain
the most direct information on the GPDs from exclusive reactions.            

\noindent 
(1) Approximations can be made for certain kinematics allowing 
to directly 
extract individual GPDs from asymmetry data~\cite{nowak}. For example, the 
GPD $H$ may be extracted from the beam asymmetry measurements for $x_B < 0.25$ 
values. Similar approximations may be possible for the GPD $\tilde{H}$ in 
target asymmetry measurements.      

\noindent
(2) Fits of GPD parametrizations to large sets of data may be employed, 
in which constraints from elastic form 
factors, meson distribution amplitudes, forward parton distributions, 
and polynomiality conditions are imposed~\cite{freund}.

\noindent
(3) A new technique, not unrelated to the focus of this workshop, 
has recently been proposed~\cite{polyakov} that make use of
partial wave analysis techniques where the GPDs are expanded in infinite sums 
over t-channel exchanges. The convergence of such a procedure, however 
remains to be explored. 

\section{``Tomographic'' Images of the Nucleon.} 

Recently it was found~\cite{burkardt,belitsky2,ralston} that knowledge of the $x$ 
and $t$ dependence of 
GPDs for specific quark flavors provides the basis for a visualization  
of the proton's  quark content in the transverse plane, in some analogy 
to the way images of macroscopic objects can be assembled in tomography. 
Using models for GPDs the quark density distribution and the quark spin 
distribution have been studied. A strong 
dependence of the quark density with increasing value of the longitudinal 
momentum fraction $x$ is found. These studies also show a very strong 
spin-flavor polarization between $u$ quarks  and $d$ quarks 
if the proton is polarized in the transverse plane. It is
found that the $u$ and the $d$ 
quark spin distribution are spatially separated from each other especially 
at high $x$. This regime can be accessed in deeply 
exclusive processes at currently available energies, and will be 
fully accessible after the JLab Upgrade.

In summary, GPDs uniquely connect the charge and current distribtions of the nucleon 
with the forward quark distributions measured in DIS. Recent results 
demonstrate the applicability of the GPD framework at currently 
achievable values of $Q^2$ for DVCS and possibly for vector meson 
production. 
There is a program underway at JLab to study the DVCS 
process at 6~GeV. A broad program of DVCS and DVMP has been 
proposed for the 12~GeV Upgrade. The  results of such a program 
will provide new insights into 
the internal dynamics of the nucleon unimaginable just five years 
ago.


\begin{thebibliography}{200}

\bibitem{mueller} D. M\"uller et al., Fortschr.Phys.42,101 (1994)
\bibitem{xji1} X. Ji, Phys.Rev.D55,7114 (1997)
\bibitem{radyush} A. Radyushkin, Phys.Lett.B380.417 (1996), Phys.Rev.D56,5524 (1997)
\bibitem{belitsky1} For details see: A. Belitsky, talk at this conference
\bibitem{burkardt} M. Burkardt, Nucl.Phys.A711(2002)127
\bibitem{belitsky2} A. Belitsky and D. M\"uller, Nucl.Phys.A711(2002)118
\bibitem{ralston}J.P. Ralston and B. Pire, Phys.Rev.D66, 111501 (2002) 
\bibitem{xji} X. Ji, Phys.Rev.Lett.78,610(1997)
\bibitem{dvcsclas} S. Stepanyan, et al., Phys.Rev.Lett.87,182002-1(2001)
\bibitem{dvcshermes} A. Airapetian, et al., Phys.Rev.Lett.87,182001 (2001)
\bibitem{mvdh} K. Goeke, M. V. Polyakov, M. Vanderhaeghen, 
Prog. Part. Nucl. Phys. 47 (2001) 401 
\bibitem{belitsky3} A. Belitsky, D. M\"uller, and D. Kirchner, 
Nucl.Phys.B629(2002)323.
\bibitem{dvcsh1} C. Adloff, et. al., Phys. Lett.B517, 47(2001)
\bibitem{dvcszeus} P.R. Saull [ZEUS collaboration] arXiv:hep-ex/0003030
\bibitem{freund} A. Freund, M. McDermott, and M. Strikman, arXiv:hep-ph/0208160
\bibitem{garcon} M. Garcon, M. Guidal, E. Smith (spokespersons), Experiment E-99-105
\bibitem{guidal} M. Guidal, private communications. 
\bibitem{hallb} V. Burkert, L. Elouadrhiri, M. Garcon, S. Stepanyan (spokespersons), 
Experiment E-01-113.
\bibitem{halla} Y. Roblin and F. Sabati\'e (spokespersons), 
JLab Experiment E-00-110.   
\bibitem{nowak} V.A. Korotkov and W.D. Nowak, Eur.Phys.J.C23:455-461,2002 
\bibitem{polyakov} M.V. Polyakov and  A.G. Shuvaev, arXiv:hep-ph/0207153  
\end{thebibliography}
\end{document}